\documentclass[prl,twocolumn,showpacs,amsmath]{revtex4}

\usepackage{bbm}
  \newcommand\One{\mathbbm{1}}          
\usepackage{slashed}                    
\usepackage{feynmp}                     
\usepackage{feynmphack}                 
\usepackage{color}

\DeclareMathOperator\Tr{Tr}
\newcommand\Mathstrut{\vphantom{\big(}}
\newcommand\twist[1]{\mbox{twist-#1}}
\newcommand\Twist[1]{\mbox{Twist-#1}}
\begin{document}

\title{Colour Modification of Effective T-odd Distributions}

\author{\firstname{Philip G.}~\surname{Ratcliffe}}
\email{philip.ratcliffe@uninsubria.it}
\affiliation{%
  Dipartimento di Fisica e Matematica, Universit\`{a} degli Studi dell'Insubria,
  via Valleggio 11, 22100 Como, Italy
}
\affiliation{%
  Istituto Nazionale di Fisica Nucleare---sezione di Milano--Bicocca,
  piazza della Scienza 3, 20126 Milano, Italy
}

\author{\firstname{Oleg V.}~\surname{Teryaev}}
\email{teryaev@thsun1.jinr.ru}
\affiliation{%
  Bogoliubov Laboratory of Theoretical Physics, Joint Institute for Nuclear
  Research, Dubna 141980, Russia
}

\date{\today}

\begin {abstract}
We show that soft-gluon \twist3 contributions to single-spin asymmetries (SSA) in hard processes may be expressed in the form of \emph{effective} T-odd Sivers distributions, whose signs and scales are modified by process-dependent colour factors. We thus prove that the Sivers mechanism may also be applied at large transverse momenta. We stress that \twist3 SSA in semi-inclusive deeply inelastic scattering and  Drell--Yan processes are suppressed by transverse momentum rather than a virtual-photon momentum transfer and thus na\"{\i}vely correspond to twist two at the hadronic level. More rigorously, the transverse-momentum weighted averages of the Sivers function correspond to increasing twist ($3$, $5$, $7$, \dots) while the full $k_T^{}$-dependent Sivers function (just as other transverse-momentum dependent distribution and fragmentation functions) corresponds to a resummed infinite tower of higher twists.
\end{abstract}


\maketitle

\section {Introduction}

Single-spin asymmetries (SSA) represent one of the most subtle and intriguing effects in QCD. In the simplest inclusive processes parity conservation requires a transversely polarised beam or target. Inasmuch as the transverse polarisation component is not enhanced by Lorentz boosts, one immediately encounters the necessity of describing \twist3 effects. This can be achieved via use of either local \cite{Shuryak:1981pi} or non-local \cite{Bukhvostov:1984as,Efremov:1983eb,Ratcliffe:1985mp,Balitsky:1987bk} operators.

The latter approach also permits the description of the imaginary phases required to produce T-odd effects, such as SSA. These phases mimic true T(CP) violation (see \emph{e.g.} \cite{Teryaev:2000pi}) and allow T-odd effects in a T-conserving theory, such as QCD. The phases emerging from gluon loops describing initial- and final-state interactions (ISI and FSI) in hard subprocesses are suppressed by powers of light-quark masses and the QCD coupling constant \cite{Kane:1978nd}. However, deeper analysis \cite{Efremov:1981sh} shows that quark masses should be substituted by hadronic mass scales. Moreover, ISI and FSI between the hard and soft regions of QCD factorisation, which is just the physical picture corresponding to twist three, lead to SSA free of both suppression factors \cite{Efremov:1984ip}. The imaginary phase is generated by gluon correlations with soft quarks; the situation when instead the gluon is soft was also considered later~\cite{Qiu:1991pp}.

An alternative description of SSA effects is provided by a T-odd transverse-momentum dependent (TMD) distribution function, first introduced by Sivers \cite{Sivers:1989cc}. As soon as there is no kinematical variable whose cut produces an imaginary phase in the hadron--parton transition amplitude, this may simply become an effective function \cite{Teryaev:2000pi}, so that the phase also emerges owing to the ISI and FSI involving hard subprocess. The first case of the appearance of an effective T-odd distribution was found \cite{Boer:1997bw} for soft-gluon SSA in the Drell--Yan (DY) process integrated over transverse momenta \cite{Hammon:1996pw}. It was later identified \cite{Boer:2003cm} with the first moment of the Sivers function, which plays a special role in what follows. The role of FSI between the hard and soft regions of semi-inclusive deeply inelastic scattering (SIDIS) was clearly revealed in the model of Brodsky, Hwang and Schmidt \cite{Brodsky:2002cx}, where it was interpreted \cite{Collins:2002kn.x} as a manifestation of the Sivers function. The crucial role of hard processes in defining this function was made manifest by the discovery of a sign difference between SIDIS and DY.

This is all qualitatively similar to earlier findings \cite{Teryaev:2002wf} in the \twist3 case. However, the apparent difference between the FSI arising in \twist3 interactions is the absence of true power suppression. The situation is, though, even more peculiar. In the hard Abelian process of semi-inclusive production of a real photon by a deeply virtual photon (SIDVCS, the semi-inclusive counterpart of the well-known DVCS \cite{Ji:1996nm} process) an overall suppression as $Mp_T^{}/Q^2$ was shown \cite{Teryaev:2005bp} to be compensated by a gluonic pole in the quark--gluon correlator, which is approached at low $p_T^{}\ll{}Q$ as the gluon momentum fraction is defined by kinematics $x_g\sim{}p_T^2/Q^2$, indicating the possibility to obtain unsuppressed (in $Q$) \twist3 effects. Similar conclusions that the Sivers function and gluonic poles describe similar physics for different $p_T^{}$ have been reached within the framework of a general proof~\cite{Ji:2006ub}.

These analyses imply a picture in which the Sivers function is limited to the low-$p_T^{}$ region, where a special type of factorisation \cite{Collins:2004nx} is assumed valid and either the continuation of the \twist3 result to lower $p_T^{}$ \cite{Teryaev:2005bp} or matching \cite{Ji:2006ub} of high- and low-$p_T^{}$ results is adopted.

Here we suggest a different, complementary approach, to apply the Sivers function at \emph{high} $p_T^{}$. This is of special importance for hadronic processes where $p_T^{}$ is the only hard scale. We present general quantitative relations between the Sivers function and gluonic poles, using master formul{\ae} \cite{Koike:2006qv.x} for the latter, and find that besides the sign there are important process-dependent colour factors (\emph{cf}.\ \cite{Bomhof:2006np,Bomhof:2007su.x}, where such colour factors were calculated by considering gauge links) modifying the Sivers function and underlining its effective nature.

We also provide a general proof of the non-suppression in $Q^2$ of the \twist3 SSA in question. We show that the various transverse moments of the Sivers function correspond to increasing twist ($3$, $5$, $7$, \dots), while the entire TMD function corresponds to a resummed infinite tower of higher twists. We also perform the first rigorous application of the Sivers function to SSA at large $p_T^{}$ and discuss some phenomenological consequences

\section {From the Sivers function\break to gluonic poles}

To prove the relation between twist three and the Sivers function we shall not attempt to obtain the latter as some special limit \cite{Teryaev:2005bp,Ji:2006ub} of a \twist3 contribution, but instead transform some approximation of it to the form \cite{Koike:2006qv.x} appearing in the \twist3 calculation.

We shall start with the following factorised formula involving the Sivers function
\begin{equation}
  \label{SRS}
  d\Delta\sigma \sim
  \int \! d^2k_T^{} dx \, f_{\text{S}}^{} (x,k_T^{}) \,
  \Tr\!\left[\Mathstrut \gamma_\rho H(xP,k_T^{}) \right]
  \epsilon^{\rho s P k_T^{}},
\end{equation}
where other (unpolarised and collinear) distribution or fragmentation functions should normally also be present. This leads to the dependence of $H$ on the respective transverse momenta. The presence of only one TMD distribution function makes a notable difference with respect to standard TMD factorisation \cite{Collins:2004nx, Ji:2004wu}, where all non-perturbative ingredients are transverse-momentum dependent.
In the latter case a similar expression is only assumed to be applicable at low transverse momenta.

We shall now prove that at large transverse momenta it is, instead, related to the contribution of gluonic poles in \twist3 factorisation. This will allow us to justify the applicability of the Sivers function in an extended kinematical region and to estimate the accuracy of such an application, depending on the degree to which the gluonic-pole contribution may be considered dominant. To achieve this goal,
we expand the subprocess coefficient function $H$ in powers of $k_T^{}$, retaining only the first non-vanishing term:
\begin{equation}
  \label{SRS1}
  d\Delta\sigma \sim
  \int \!\! d^2k_T^{} dx \, f_{\text{S}}^{}(x,k_T^{})
  \Tr\!\left[
    \gamma_\rho \frac{\partial H(xP,k_T^{})}{\partial k_T^\alpha}
  \right]_{k_T^{}{=}0}
  \mkern-35mu
  k_T^\alpha \, \epsilon^{\rho s P k_T^{}}\!.
\end{equation}
This is a natural step in selecting \twist3 terms \cite{Efremov:1983eb,Ratcliffe:1985mp}, justified by the expected rapid decrease of the Sivers function with respect to any hard scale determining the $k_T^{}$ dependence of the hard kernel, and is a crucial element of our proof.

The $k_T^{}$ integration now only includes soft parts and will thus be expressed via moments of the Sivers function, which may otherwise appear due to particular definitions of some measurable asymmetries \cite{Bomhof:2007su.x}. Let us first average over the directions of $k_T^{}$ using the standard relation
\begin{equation}
  \left<\Mathstrut k_T^\mu k_T^\nu \right> =
  -\tfrac{1}{2} g_T^{\mu\nu} \left< k_T^2 \right>,
\end{equation}
where $g_T^{\mu \nu}\equiv{}g^{\mu\nu}-P^\mu n^\nu-n^\mu P^\nu$ is defined with respect to the same light-cone vectors $P$ and $n$ ($P{\cdot}n=1$) that define the direction of $k_T^{}$, thus $P{\cdot}k_T^{}=0=n{\cdot}k_T^{}$. On substituting this expression into (\ref{SRS1}), one obtains
\begin{multline}
  \label{SRS2}
  d\Delta\sigma \sim
  -M \int \! dx \, f_{\text{S}}^{(1)}(x)
  \Tr\!\left[
    \gamma_\rho \frac{\partial H(xP,k_T^{})}{\partial k_T^\alpha}
  \right]_{k_T^{}{=}0}
\\
  \null \times
  \left( \Mathstrut
    \epsilon^{\rho s P \alpha} - P^\alpha \epsilon^{\rho s P n}
  \right),
\end{multline}
where
\begin{equation}
  \label{mom}
  f_{\text{S}}^{(1)}(x) = \int \! d^2k_T^{} \, f_{\text{S}}^{}(x,k_T^{}) \, \frac{k_T^2}{2M^2}.
\end{equation}

The final step now exploits the following kinematic identity~\cite{Efremov:1983eb}:
\begin{equation}
  \label{ir}
  \epsilon^{\rho s P \alpha} =
  P^\alpha \epsilon^{\rho s P n} - P^\rho \epsilon^{\alpha s P n},
\end{equation}
which results from the vanishing of any totally antisymmetric fifth-rank tensor in four-dimensional space. This then allows the two terms in (\ref{SRS2}) to be recombined:
\begin{equation}
  d\Delta\sigma \sim
  M \int \! dx \, f_{\text{S}}^{(1)}(x) \,
  \Tr\!\left[
    \slashed{P} \frac{\partial H(xP,k_T^{})}{\partial k_T^\alpha}
  \right]_{k_T^{}{=}0}
  \epsilon^{\alpha s P n}.
\end{equation}
Note that the above-mentioned dependence of $H$ on the momenta of other particles, described by collinear distribution and fragmentation functions, is now crucial, as only these momenta may carry the index $\alpha$ while the derivative is calculated. If other detected particles are absent, as in the case of DIS, the asymmetry is, of course, identically zero.
The key observation now is that this expression exactly coincides with the recently obtained master formula \cite{Koike:2006qv.x} for the contribution of \twist3 gluonic poles in high-$p_T^{}$ processes. The Sivers distribution can then be identified with the gluonic pole strength $T(x,x)$ multiplied by a process-dependent colour factor. In turn, the sign of the Sivers function is fixed according as to which of the ISI or FSI is relevant:
\begin{equation}
  \label{ST}
  f_{\text{S}}^{(1)}(x) = \sum_i C_i
  \frac{1}{2M} T(x,x),
\end{equation}
where
$C_i$ is a relative colour factor, defined with respect to an Abelian subprocess (say SIDVCS discussed above, where it is just $C_F^{}$), which is naturally absorbed into the definition of the quark--gluon correlator \cite{Efremov:1983eb}. As we shall discuss below, this is also the factor appearing in low-$p_T^{}$ SIDIS and DY at the Born level.

The relation established is one of the principal results of this paper. Note first that it clarifies the relation between \twist3 and \twist2 effects discussed above. Indeed, the mass parameter in the numerator is compensated by a kinematical variable, which is produced by taking the derivative of the hard kernel with respect to transverse momentum. If it depends on both the transverse momentum $p_T^{}$ and the (larger) virtual photon momentum $Q^2$, then the terms with $1/p_T^2$ are dominant. Thus, the result is not suppressed as $1/Q^2$ and is na\"{\i}vely of leading twist.

On the other hand, the second moment of the Sivers function enters the original expression (\ref{SRS}) with a factor $M$ instead of $1/M$, indicating its \twist3 nature. This may be seen immediately by defining the Sivers function in coordinate (impact-parameter) space, in a manner similarly to earlier discussions \cite{Teryaev:2004df} of the Collins fragmentation function:
\begin{equation}
  \label{imp}
  \left< P,s | \psi(0) \gamma_\rho \psi(z) | P,s \right>
  \sim
  M \epsilon^{\rho s P z} \int \! dx \; e^{izx} f_{\text{S}}^{(1)}(x).
\end{equation}
Note too that higher Sivers-function moments enter with higher derivatives of the coefficient function and therefore correspond to higher twist ($5$, $7$, $9$, \dots). The entire $k_T^{}$-dependent Sivers function thus corresponds to a resummed infinite tower of higher twists. This property has also been studied in coordinate space \cite{Teryaev:2004df}, where $k_T^{}$-dependent functions represent a complete similarity with non-local quark condensates. The latter manifest a similar resummation of an infinite tower of higher twists (see \emph{e.g.} \cite{Bakulev:2002hk} and refs.\  therein), but for vacuum rather than hadronic matrix elements.
Note also the difference between the result obtained here and that of \cite{Ji:2006ub}: in the latter case the \twist3 contribution at large transverse momenta $p_T^{}$ is expressed in terms of a \emph{perturbative} Sivers function,
which is the source  of large $p_T^{}$.
In contrast, our result implies the use of an ordinary, non-perturbative, Sivers function.

\section{%
  Colour factors and \break
  the transition from large \break
  to small transverse momenta
}

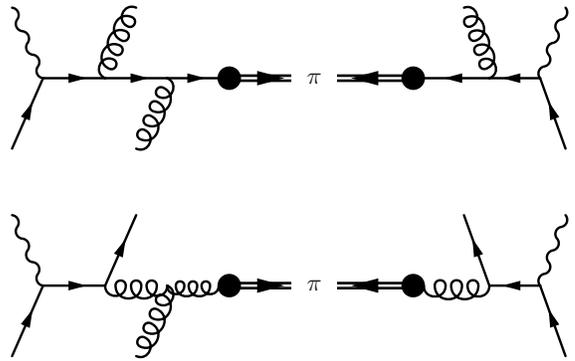
\begin{figure}[hbt]
  \centering
  \begin{fmffile}{fmf}
    \fmfset{arrow_len}{2.5mm}
    \fmfset{curly_len}{2.2mm}
    \vspace{2ex}
    \begin{fmfgraph*}(55,25)
      \fmfleft{i1,i2}
      \fmfright{d1,d2,d3}
      \fmftop{o2}
      \fmfbottom{o1}
      \fmf{phantom}{i1,v1,v2,o1}
      \fmf{phantom}{i2,v1}
      \fmf{phantom}{v2,o2}
      \fmffreeze
      \fmf{fermion}{i1,v1,v2,v3,v4}
      \fmf{dbl_plain_arrow}{v4,d2}
      \fmffreeze
      \fmf{photon}{v1,i2}
      \fmf{gluon}{o2,v2}
      \fmf{gluon}{o1,v3}
      \fmfv{d.sh=circle,d.fi=full,d.si=4thick}{v4}
      \fmfv{l=$\pi$}{d2}
    \end{fmfgraph*}
    \hspace{1.2em}
    \begin{fmfgraph}(45,25)
      \fmfright{i1,i2}
      \fmfleft{d1,d2,d3}
      \fmftop{o2}
      \fmfbottom{o1}
      \fmf{phantom}{i1,v1,v2,o1}
      \fmf{phantom}{i2,v1}
      \fmf{phantom}{v2,o2}
      \fmffreeze
      \fmf{fermion}{i1,v1,v2,v4}
      \fmf{dbl_plain_arrow}{v4,d2}
      \fmffreeze
      \fmf{photon}{v1,i2}
      \fmf{gluon}{v2,o2}
      \fmfv{d.sh=circle,d.fi=full,d.si=4thick}{v4}
    \end{fmfgraph}
    \\[6ex]
    \begin{fmfgraph*}(55,25)
      \fmfleft{i1,i2}
      \fmfright{d1,d2,d3}
      \fmftop{o2}
      \fmfbottom{o1}
      \fmf{phantom}{i1,v1,v2,o1}
      \fmf{phantom}{i2,v1}
      \fmf{phantom}{v2,o2}
      \fmffreeze
      \fmf{fermion}{i1,v1,v2}
      \fmf{gluon}{v2,v3,v4}
      \fmf{dbl_plain_arrow}{v4,d2}
      \fmffreeze
      \fmf{photon}{v1,i2}
      \fmf{fermion}{v2,o2}
      \fmf{gluon}{o1,v3}
      \fmfv{d.sh=circle,d.fi=full,d.si=4thick}{v4}
      \fmfv{l=$\pi$}{d2}
    \end{fmfgraph*}
    \hspace{1.2em}
    \begin{fmfgraph}(45,25)
      \fmfright{i1,i2}
      \fmfleft{d1,d2,d3}
      \fmftop{o2}
      \fmfbottom{o1}
      \fmf{phantom}{i1,v1,v2,o1}
      \fmf{phantom}{i2,v1}
      \fmf{phantom}{v2,o2}
      \fmffreeze
      \fmf{fermion}{i1,v1,v2}
      \fmf{gluon}{v4,v2}
      \fmf{dbl_plain_arrow}{v4,d2}
      \fmffreeze
      \fmf{photon}{v1,i2}
      \fmf{fermion}{v2,o2}
      \fmfv{d.sh=circle,d.fi=full,d.si=4thick}{v4}
    \end{fmfgraph}
    \\[3ex]
  \end{fmffile}
  \caption{%
    \Twist3 SIDIS pion production via (top) quark and (bottom) gluon
    fragmentation.
  }
  \label{fig:SIDIS-3}
\end{figure}

Let us consider some particular applications of this relation, starting with high-$p_T^{}$ SIDIS. In this case there are only final-state interactions, while the colour factors differ for mesons produced in fragmentation of quarks ($-1/2N_c$; see Fig.~\ref{fig:SIDIS-3}, top) or gluons ($N_c/2$; see Fig.~\ref{fig:SIDIS-3}, bottom).
This shows that there is a specific enhancement in the latter, which is of special importance for $K^-$ mesons.

Matching of the large- and small-$p_T^{}$ descriptions is an intriguing question as there is no sharp border. Clarification may be obtained by comparison with the perturbative Sivers-function calculation~\cite{Ji:2006ub}: there is a hard (in addition to soft) gluonic-pole contribution (when the gluon carries a finite momentum fraction) restoring the correct ``Abelian'' colour factor $C_F^{}$. Moreover, the entire surviving contribution is just due to hard poles.

It is interesting to consider the possibility of describing these effects within the framework of a model, such as that of \cite{Brodsky:2002cx}. To do so, one may modify it by considering the emission of an extra hard gluon, in order to allow for high $p_T^{}$. Moreover, to distinguish the colour factors, one should attribute some colour-charge exchange to both the FSI and to the emission of the extra hard gluon. Another possibility may be to consider the \emph{electric} instead of colour charge, following the recent idea of Collins and Qiu~\cite{Collins:2007nk}.

Generalising their approach, one may consider the emission of a hard charged gluon, balancing the large transverse momentum. In this case, the hard gluonic pole leading to the Abelian colour factor (or to the electric-charge factor corresponding to the charge of the quark from the polarised target in the approach of \cite{Collins:2007nk}) in SIDIS corresponds to FSI that occur \emph{before} emission of this extra gluon.  At the same time, the description of the \twist3 mechanism at high $p_T^{}$ in terms of an effective Sivers function related to soft poles emerges when the FSI happen after emission of the extra gluon, as in Fig.~\ref{fig:SIDIS-3} (top). Therefore, even in SIDIS factorisation, in the sense of \cite{Collins:2007nk}, is broken by the emission of charged gluons; that is, in the sense of the present paper, it is modified by the (colour) charge factor.

We note the similarity with the SIDVCS calculation~\cite{Teryaev:2005bp}, where the entire contribution associated with the Sivers function is also due to hard poles, since for finite $p_T^{}$ there are no soft poles at all. With decreasing $p_T^{}$, the hard pole becomes soft and produces the Sivers function. We therefore expect that, with decreasing $p_T^{}$, the increasing contributions of softening hard poles should smoothly alter the colour factor of quark fragmentation contributions to $C_F^{}$ while the gluonic contributions disappear, both effects occurring at typical hadronic scales. It would be challenging to verify this theoretical picture by experimental observation. We might also note that the Abelian colour factor is recovered if the fragmentation probabilities from quark and gluon jets coincide. This is natural, since for small $p_T^{}$ there is no way to distinguish or resolve quarks and collinearly emitted gluons.

To experimentally verify such a picture, it would be of major importance to distinguish between mesons originating from either quark or gluon fragmentation at large $p_T^{}$. While a complete separation is impossible, there are methods that can help. Firstly, one may use jet shape, which differs for quark and gluon jets owing to the different spins of the fragmenting objects. This difference in spin can also be seen in the tensor polarisation of vector mesons \cite{Schafer:1999am}. However, most promising would seem to be exploration of the different $z$-dependences in quark and gluon fragmentation functions. The faster decrease of the latter should result in dramatic variations of SSA, so that at low $z$ gluon fragmentation would be dominant with a colour factor $N_c/2$, while at large $z$ one would expect a sign change and transition to quark fragmentation with a factor~$-1/2N_c$.

Note that at high $p_T^{}$ there is thus a simple and direct connection between the Sivers functions for SIDIS and DY processes. Indeed, the relation is now
\begin{equation}
  f_{\text{S}}^{\text{SIDIS},i} = -f_{\text{S}}^{\text{DY},i} \quad (i=q,g),
\end{equation}
which holds separately for quark and gluon contributions. In DY this entails dilepton production by quarks or gluons from an unpolarised hadron, while in SIDIS it corresponds to meson production from quark or gluon fragmentation. The observable asymmetries are therefore proportional to distribution and fragmentation functions respectively. This relation is also very interesting to test experimentally, especially since the (unpolarised) gluon distribution function is much better known than that for fragmentation.

Let us now turn to hadronic processes, starting with the simplest: direct-photon production. There are only initial-state interactions with gluons, resulting in the very simple relation
\begin{equation}
  F_{\text{S}}^{hh\to\gamma X} = \frac{N_c}{2} \, f_{\text{S}}^{\text{DY}}.
\end{equation}
Exploration of this process in various kinematical regions can provide information on the gluon Sivers function~\cite{Schmidt:2005gv}. There is little doubt that this is also an effective function, related to the three-gluon correlators considered earlier in relation to pion SSA \cite{Ji:1992eu} and the DIS structure function $g_2$ \cite{Belitsky:1995vd,Belitsky:2000pb}. The generalisation of our approach to the case of three-gluon correlators is therefore an important task. Consideration of quark--gluon processes is more complicated; a list of colour factors relevant for \twist3 subprocesses may be found in \cite{Bomhof:2006np,Bomhof:2007su.x}. Let us only mention that FSI for pions produced in quark fragmentation may be reexpressed in a manifestly gauge-invariant manner via the summation formula
\begin{equation}
  t^a S \, t^a = -\frac{1}{2N_c} \, S + \frac{1}{2} \One \Tr{S}.
\end{equation}
The first term corresponds to the usual Sivers function \cite{Anselmino:2004ky} with colour factor $-1/2N_c$ and the second to the Abelian Compton subprocess, with $s$- and $u$-channel diagrams contributing with the same factors while the $t$-channel is absent. Both terms are separately gauge invariant.

Such a dramatic modifications of hadronic processes may be considered as another  way of describing what, in the model approach \cite{Collins:2007nk} discussed above, was termed the violation of factorisation in hadronic processes. The calculation was performed there in an Abelian model similar to that of~\cite{Brodsky:2002cx}.

The main point of the argument in \cite{Collins:2007nk} is the proportionality of the contribution of the Sivers function to the electric charge of the quark from the other (unpolarised) hadron. This observation has a direct counterpart in our approach. The colour factor is defined by the colour charge of the parton participating in the ISI and FSI. This charge is, generally speaking, independent of the properties of the polarised hadron emitting the gluon that participates in the ISI and FSI and, in this sense, breaks factorisation. SIDIS and DY processes at low $p_T^{}$ are exceptional: the colour charge of the quark participating in the FSI in SIDIS is the same as that of the quark originally emitted by polarised hadron. By the same token, the colour charge of the antiquark participating in the ISI in DY processes at low $p_T^{}$ is just the opposite, which explains the Collins sign rule. At the same time, the emission of a hard gluon changes these colour charges in high-$p_T^{}$ SIDIS, DY processes and, needless to say, other hadronic processes. This modification of colour charge causes a colour modification of the effective Sivers function.

\section{Discussion and Conclusions}

We have suggested and proved here a method of applying the Sivers distribution at large transverse momenta. We have shown that the Sivers function is, in effect, none other than an expression of the contribution of gluonic poles. It is therefore process dependent and this dependence includes, besides the sign related to the ISI and FSI responsible for the imaginary phase, a colour factor.

Such a picture is complementary to that considered previously, in which matching between the Sivers function and \twist3 matrix elements occurred in the region where, strictly speaking, factorisation formul{\ae} were not valid. The matching between various $p_T^{}$ regions now takes the form of a $p_T^{}$-dependent colour factor. We have studied this dependence in SIDIS, where it is a sort of colour separation, similar to the phenomenon of colour transparency. That is, at low $p_T^{}$ one cannot distinguish between mesons originating from quark and gluon fragmentation. As soon as one is able to do so at larger $p_T^{}$, they enter the Sivers asymmetry with dramatically different colour factors, in both sign and scale.

This complementary method of establishing a relation between the Sivers function and \twist3 matrix elements lends support to the possibility of global fits of Sivers functions~\cite{Teryaev:2005bp}, including lepton--hadron and hadron--hadron processes, as well as DIS, where twist three also contributes.

We have shown that transverse moments of the Sivers function correspond to increasing twists, starting at three. The entire function corresponds to a resummed tower of twists, which is just the object to be considered at low~$p_T^{}$.

Our result is especially relevant for those hadronic processes in which $p_T^{}$ is the only hard scale. We have proved that direct-photon production is described by a colour factor $N_c/2$ and have suggested a method to rearrange the colour factors in a manifestly gauge-invariant way.

\begin{acknowledgments}
We are indebted to Anatoly V. Efremov and Andreas Metz for useful discussions. O.V.T. is grateful to the Cariplo Science Foundation for support during his stay at the University of Insubria in Como and to the Department of Physics and Mathematics of the University for kind hospitality. This work was partially supported by the Deutsche Forschungsgemeinschaft, grant 436 RUS 113/881/0, the Russian Foundation for Basic Research (Grant 03-02-16816) and the Russian Federation Ministry of Education and Science (Grant MIREA 2.2.2.2.6546).
\end{acknowledgments}


\end{document}